\documentstyle[11pt,aaspp,epsf]{article}

\def\refitem #1! #2! #3! #4;{\hang\noindent
    \hangindent 20pt\rm #1, \rm #2, \rm #3, \rm #4.\par}
\def\bookref{\par\noindent\hangindent 20pt}

\def\etal   {{\sl et~al.}}
\def\wisk#1{\ifmmode{#1}\else{$#1$}\fi}
\def\percm  {\wisk{{\rm cm}^{-1}}}

\def\percmcub{\wisk{{\rm cm}^{-3}}}

\def\gsim   {\wisk{_>\atop^{\sim}}}

\def\um     {\wisk{{\rm \mu m}}}

\def\COBE{{\sl COBE\/}}
\def\muK     {\wisk{{\rm \mu K}}}
\def\kmsMpc  {\wisk{{\rm km}\;{\rm s}^{-1}\;{\rm Mpc}^{-1}}}
\def\Amp     {\wisk{{\langle Q_{RMS}^2\rangle^{0.5}}}}

\begin{document}
\pagestyle{empty}

\noindent
\makebox[0pt][l]{
\raisebox{36pt}[0pt][0pt]{
astro-ph/9407056, UCLA-ASTRO-ELW-93-01, BCSPIN Lecture Notes}}

{\Large COBE
\footnote{
The National Aeronautics and Space Administration/Goddard Space Flight
Center (NASA/GSFC) is responsible for the design, development, and
operation of the Cosmic Background Explorer (COBE).
Scientific guidance is provided by the COBE Science Working Group.
GSFC is also responsible for the development of the analysis software and
for the production of the mission data sets.
} 
Tutorial and Recent Results}

\vspace{0.2in}

\makebox[0.5in]{}\parbox[t]{3in}
{Edward L. Wright\newline
UCLA Dept. of Astronomy\newline
Los Angeles CA 90024-1562}

\begin{abstract}
Some of the technical details involved in taking and analyzing data
from the \COBE\ are discussed, and
recent results from the FIRAS and DMR experiments are summarized.
Some of the cosmological implications of these recent data are
presented.
\end{abstract}

\section{Introduction}

The \COBE\ mission is the product of many years of work by a large team of
scientists and engineers.  Credit for all of the results presented in these
lectures must be shared with the other members of the the \COBE\ Science
Working Group: Chuck Bennett, Ed Cheng, Eli Dwek, Mike Hauser, Tom Kelsall,
John Mather, Harvey Moseley, Nancy Boggess, Rick Shafer, Bob Silverberg, 
George Smoot, Steve Meyer, Rai Weiss, Sam Gulkis, Mike Janssen, 
Dave Wilkinson, Phil Lubin, and Tom Murdock.  
Bennett \etal\ (1992a) is an excellent review of the history of the \COBE\
project and its results up to the discovery of the anisotropy by the DMR,
but not including the latest FIRAS limits on distortions.

Some of the members of this team have been working on \COBE\ since 1974, when
the proposals for what became the \COBE\ project were submitted.  I have been
working on \COBE\ since the beginning of 1978.
\COBE\ was successfully launched on 18 November 1989 from California, and
returned high quality scientific data from all three instruments for ten 
months until the liquid helium ran out.  But about 50\% of the instruments 
do not require liquid helium, and are still returning excellent scientific 
data in January 1993.  
While the \COBE\ mission has been very successful,
making the ``discovery of the century'', one must remember that this work is
based on the earlier work (in the $20^{th}$ century!) of Hubble (1929)
and Penzias and Wilson (1965), who discovered the expansion of the Universe 
and the microwave background itself.  As a consequence of these two 
discoveries, one knows that the early Universe was very hot and dense.  
When the density and
temperature are high, the photon creation and destruction rates are very high,
and are sufficient to guarantee the formation of a very good blackbody
spectrum.  Later, as the Universe expands and cools, the photon creation and
destruction rates become much slower than the expansion rate of the Universe,
which allow distortions of the spectrum to survive.  In the standard model the
time from which distortions could survive is 1 year after the Big Bang at
a redshift $z \approx 10^{6.4}$.

However, the action of the
expansion itself on a blackbody results in another blackbody with a lower
temperature.  Thus the existence of a distorted spectrum in the hot Big Bang
model requires the existence at time later than 1 year after the Big Bang
of both an energy source and an emission
mechanism that can produce photons that are now in the millimeter spectral 
range.  Conversely, a lack of distortions can be used to place limits
on any such energy source, such as decaying neutrinos, dissipation of
turbulence, etc.

At a time $3\times10^5$ years after the Big Bang, at $z \approx 10^3$, the
temperature has fallen to the point where helium and then hydrogen 
(re)combine into transparent gases.  
The electron scattering which had impeded the free
motion of the CMB photons until this epoch is removed, and the photons stream
across the Universe.  Before recombination, the radiation field at any point
was constrained to be very nearly isotropic because the rapid scattering
scrambled the directions of photons.  
The radiation field was not required
to be homogeneous, because the photons remained approximately fixed in 
comoving
co-ordinates.  After recombination, the free streaming of the photons has the
effect of averaging the intensity of the microwave background over a region
with a size equal to the horizon size.  Thus after recombination any
inhomogeneity in the microwave background spectrum is smoothed out.  
Note that this
inhomogeneity is not lost: instead, it is converted into anisotropy.  
When we
study the isotropy of the microwave background, we are looking back to the
surface of last scattering 300,000 years after the Big Bang.  
But the hot spots
and cold spots we are studying existed as inhomogeneities in the Universe
before recombination.  
Since the $7^\circ$ beam used by the DMR instrument on
\COBE\ is larger than the horizon size at recombination, these 
inhomogeneities cannot be constructed in a causal fashion during the epoch 
before recombination 
in the standard Big Bang model.   Instead, they must be installed ``just so''
in the initial conditions.  In the inflationary scenario of Guth (1980), 
these large scale structures were once smaller than the horizon size during 
the inflationary epoch, but grew to be much larger than the horizon.  Causal
physics acting $10^{-35}$~seconds after the Big Bang can produce the
large-scale inhomogeneities studied by the DMR.

\section{COBE Orbit and Attitude}

\COBE\ was launched at 0634 PST on 18 November 1989 into a sun-synchronous
orbit with an inclination of $99^\circ$ and an altitude of 900 km.
By choosing a suitable combination of inclination and altitude, the 
precession rate of the orbit can be set to follow the motion of the Sun
around the sky at 1 cycle per year.  The gravitational potential energy
of the satellite, averaged over its orbit, has an inclination dependence
due to the equatorial bulge of the Earth that is proportional to
\begin{equation}
\Delta E \propto \frac {-m \cos^2i} {a^3}
\end{equation}
where $m$ is the mass of \COBE,
which produces a torque 
$T = -\partial E/\partial i \propto 2 m a^{-3}\cos i \sin i$.
The perpendicular component of the angular momentum of the satellite in 
its orbit is proportional to
\begin{equation}
L_\perp \propto m \sin i \sqrt{a}.
\end{equation}
The precession rate is determined by $\omega_P = T/L_\perp$ which is thus
\begin{equation}
\omega_P \approx \left(-10^\circ/{\rm day}\right) \cos i (R_\oplus/a)^{3.5}
\end{equation}
where the minus sign indicates a retrograde precession.  The choice
$i = 99^\circ$, or $\cos i = -0.156$, with $a = R_\oplus + 900\;{\rm km} =
7271\;{\rm km}$, gives the proper precession rate.

Since \COBE\ was launched toward the South in the morning, 
the end of the orbit normal which points toward the Sun is at declination
$-9^\circ$.  At the June solstice, when the declination of the Sun is 
$+23^\circ$, there is a $32^\circ$ deviation from the ideal situation
where the Earth-Sun line is exactly perpendicular to the orbit.  Since
the depression of the horizon given by $\cos d = R_\oplus/a$ is $29^\circ$,
there are eclipses when \COBE\ passes over the South pole for a period of
two months centered on the June solstice.  During the same season at the
North pole the angle between the Sun and anti-Sunward limb of the Earth
is only $177^\circ$.  It is thus not possible to keep both Sunlight and
Earthshine out of the shaded cavity around the dewar.  Since the Sun
produces a much greater bolometric intensity than a thin crescent of
Earth limb, the choice to keep the Sun below the plane of the shade is
an obvious one.

\begin{figure}[tb]
\plotone{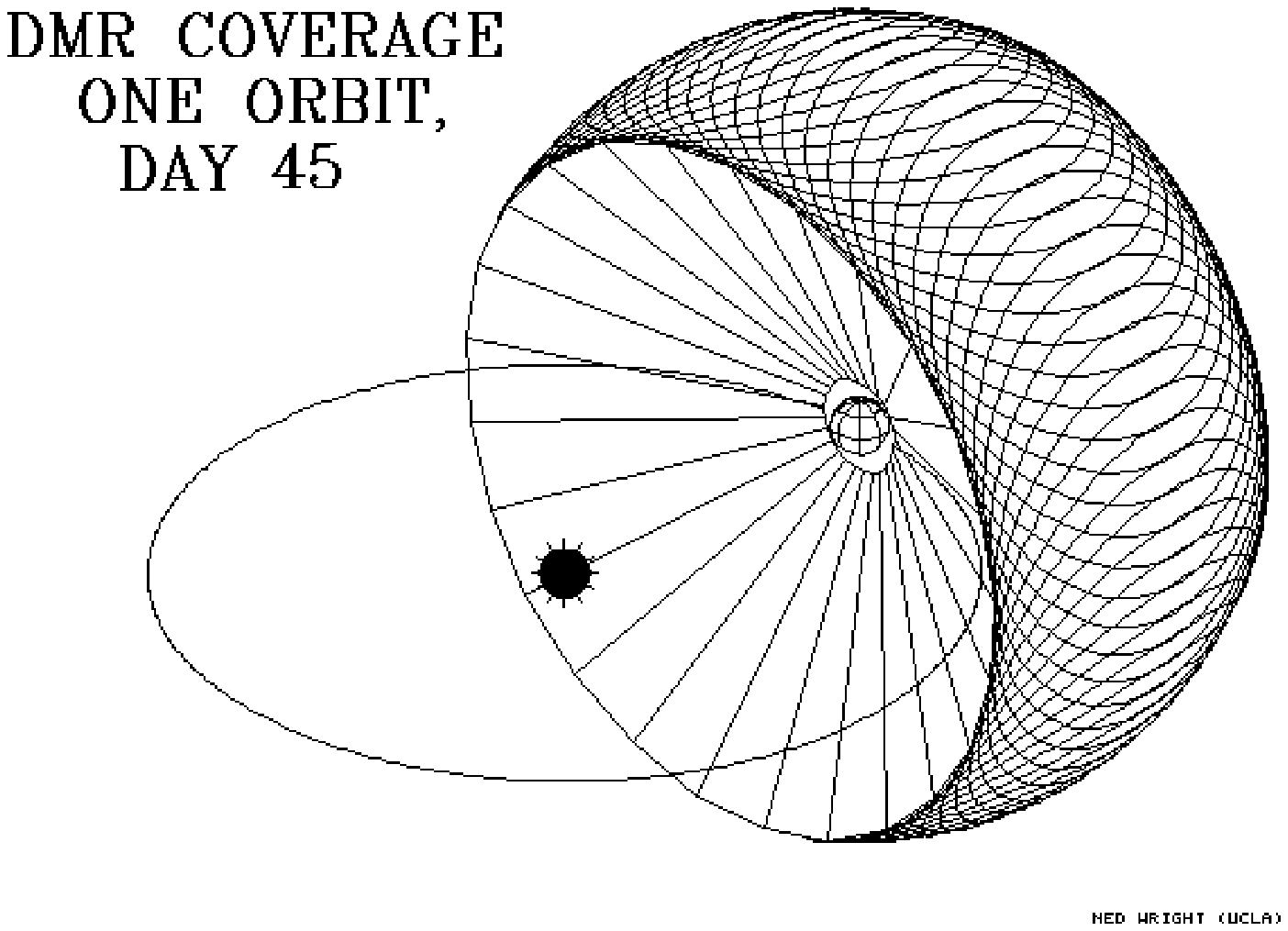}
\caption{The DMR scan pattern for one orbit covers the band from $64^\circ$
to $124^\circ$ from the Sun.}
\label{day45}
\end{figure}

The pointing of \COBE\ is normally set to maintain the angle between
the spin axis and the Sun at $94^\circ$.  This leaves the Sun
$4^\circ$ below the plane of the shade.  During eclipse seasons
this margin is shaved to $2^\circ$ to reduce the amount of Earthshine
into the shaded cavity where the instruments sit.  The angle between
the plane of the shade and the Sun is known as the ``roll'' angle.
The ``pitch'' angle describes the rotation of the spin axis around
the Earth-Sun line.  This rotation is programmed for a constant
angular rate of 1 cycle per orbit.  The nadir is set to be close
to the minus spin axis, but because the Sun can be up to $32^\circ$
from the orbit normal and the Sun is kept $4^\circ$ below the plane of
the shade, the spin axis can be up to $36^\circ$ away
from the zenith.  The uniform rate in pitch allows the nadir
to oscillate by up to $\pm 6^\circ$ relative to the spin-Sun plane
with a period of 2 cycles per orbit.  This motion is similar to the
effect of the inclination of the ecliptic on the equation of time
that can be seen in an analemma.  To keep the ``wind'' caused by the
orbital motion of \COBE\ from impinging on the cryogenic optics, the
pitch is biased backward by $6^\circ$.  The ``yaw'' motion of \COBE\ 
is its spin.  The spin period is about 73 seconds.

There are two types of instrument pointings on \COBE.  The FIRAS
points straight out along the spin axis.  As a result its field-of-view
moves only at the orbital rate of $3.6^\circ/{\rm min}$.  
Since FIRAS coadds interferograms over a collection period of about 
30 to 60 seconds, this slow scanning of the sky is essential.  
The DMR and DIRBE instruments
have fields-of-view located $30^\circ$ away from the spin axis.  As a
result they scan cycloidal paths through the sky, that cover the band
from $64^\circ$ to $124^\circ$ from the Sun.  This band contains 
50\% of the sky, and it is completely covered in a day by the DMR or
DIRBE.
Figure \ref{day45} shows the DMR scan pattern schematically.
This rapid coverage of the sky implies a rapid scan rate:
$2.5^\circ/{\rm sec}$.  The data rates are also high: the DIRBE transmits
8 readings/channel/sec, while the DMR transmits 2 readings/channel/sec.

\begin{figure}[tb]
\plotone{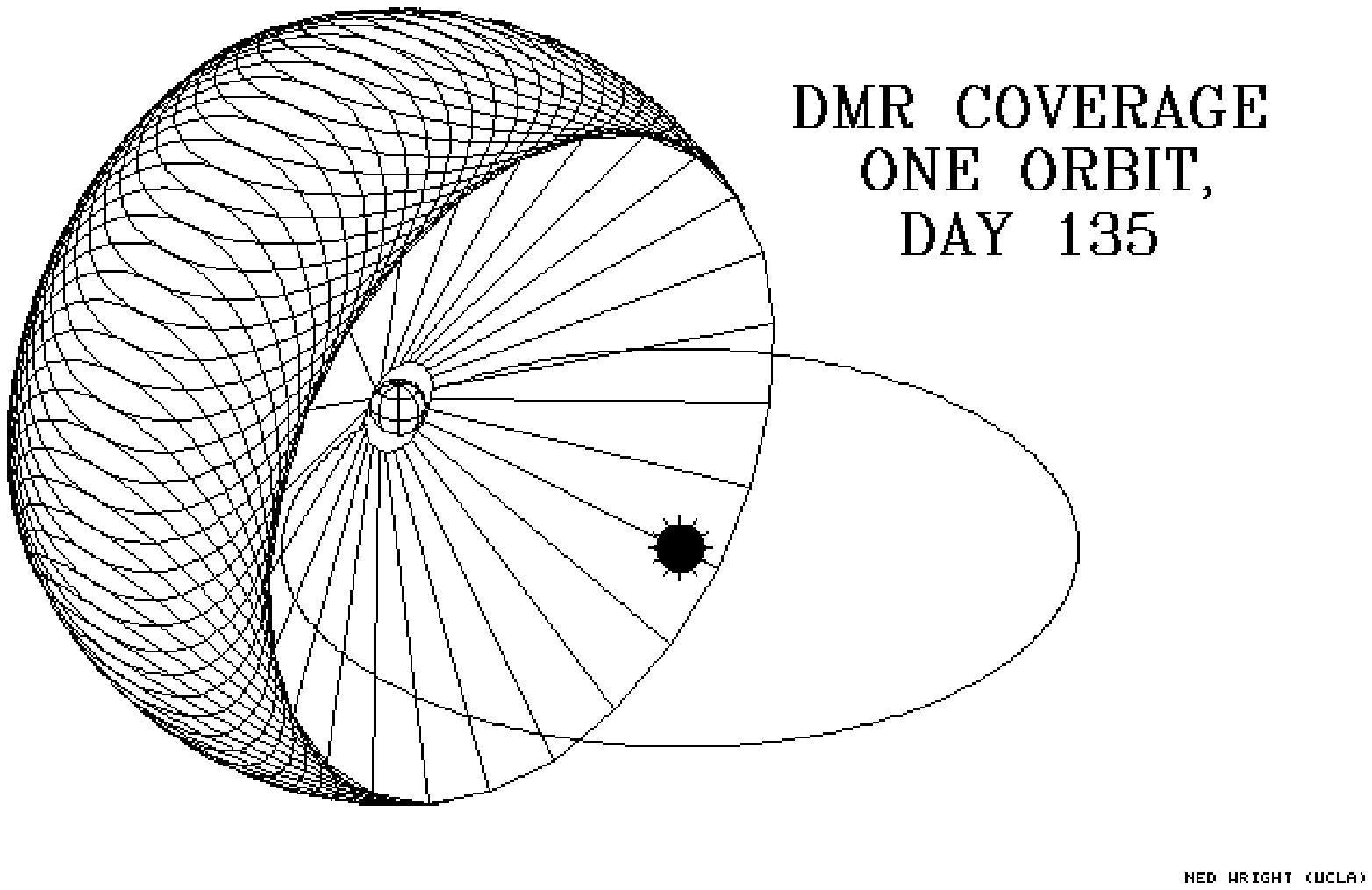}
\caption{The DMR scan pattern 3 months later with the Sun in a different
position.}
\label{day135}
\end{figure}

Over the course of a year the Sun moves around the sky, and the scan
paths of the instruments follow this motion.  The leading edge of the
wide scanned band for the DIRBE and DMR caught up to the trailing
edge of the band at start of mission after 4 months, leading to
100\% sky coverage.  
Figure \ref{day135} shows the DMR scan pattern 3 months after the
situation shown in Figure \ref{day45}.
For the FIRAS total sky coverage would in
principle be obtained after 6 months, but because of gaps in the
observations caused by calibration runs and lunar interference,
the actual sky coverage for FIRAS, defined as having the center of
the beam within a pixel, is about 90\%.  The coverage gaps
are sufficiently narrow, however, that every direction on the
sky was covered by at least the half power contour of the beam.

\section{FIRAS Measurements}

The Far InfraRed Absolute Spectrophotometer instrument on \COBE\ is
a polarizing Michelson or Martin-Puplett (1970) interferometer.  
The optical layout
is very symmetrical, and it has two inputs and two outputs.  
If the two inputs
are denoted SKY and ICAL, then the two outputs, which are denoted LEFT and
RIGHT, are given symbolically as 
\begin{eqnarray}
{\rm LEFT} & = & {\rm SKY} - {\rm ICAL}\\
{\rm RIGHT} & = & {\rm ICAL} - {\rm SKY}
\end{eqnarray}
The FIRAS has achieved its incredible sensitivity to small deviations from a
blackbody spectrum by connecting the ICAL input to an internal calibrator, a
reference blackbody that can be set to a temperature very close to the
temperature $T_\circ$ of the sky.  Thus this ``absolute'' spectrophotometer
is so successful because it is differential.
In addition, each output is further divided by a dichroic beamsplitter into a
low frequency channel (2-21 \percm) 
and a high frequency channel (23-95 \percm).
Thus there are four overall outputs.  These are labeled LL (left low)
through RH (right high).

Each output has a large composite
bolometer that senses the output power by absorbing the radiation, converting
this power into heat, and then detecting the temperature rise of a substrate
using a very small and sensitive silicon resistance thermometer.  In order to
minimize the specific heat of the bolometer, which maximizes the temperature
rise for a given input power, the substrate is made of the material with the
highest known Debye temperature: diamond.  
Diamond is quite transparent to millimeter waves, however, so an absorbing
layer is needed.  Traditionally a layer of bismuth has been used as the
absorber on composite bolometers, but the FIRAS detectors use a very thin 
layer of gold alloyed with chromium.  
The absorbing layer needs to have a surface impedance about one-half of the
377 Ohms/$\Box$ impedance of free space in order to absorb efficiently, and
this surface impedance can be obtained with a much thinner layer of gold 
rather than bismuth.  Again, this serves to minimize the specific heat of the
detector.  In order to detect the longest waves
observed by FIRAS, the diameter of the octagonal diamond substrate is quite
large: about 8 mm.  Each detector sits behind a compound parabolic
concentrator, or Winston cone, which provides a full $\pi$ steradians in the
incoming beam.  As a result, the FIRAS instrument has a very large \'etendue
of 1.5 cm$^2\;$sr.

\begin{figure}[tb]
\plotone{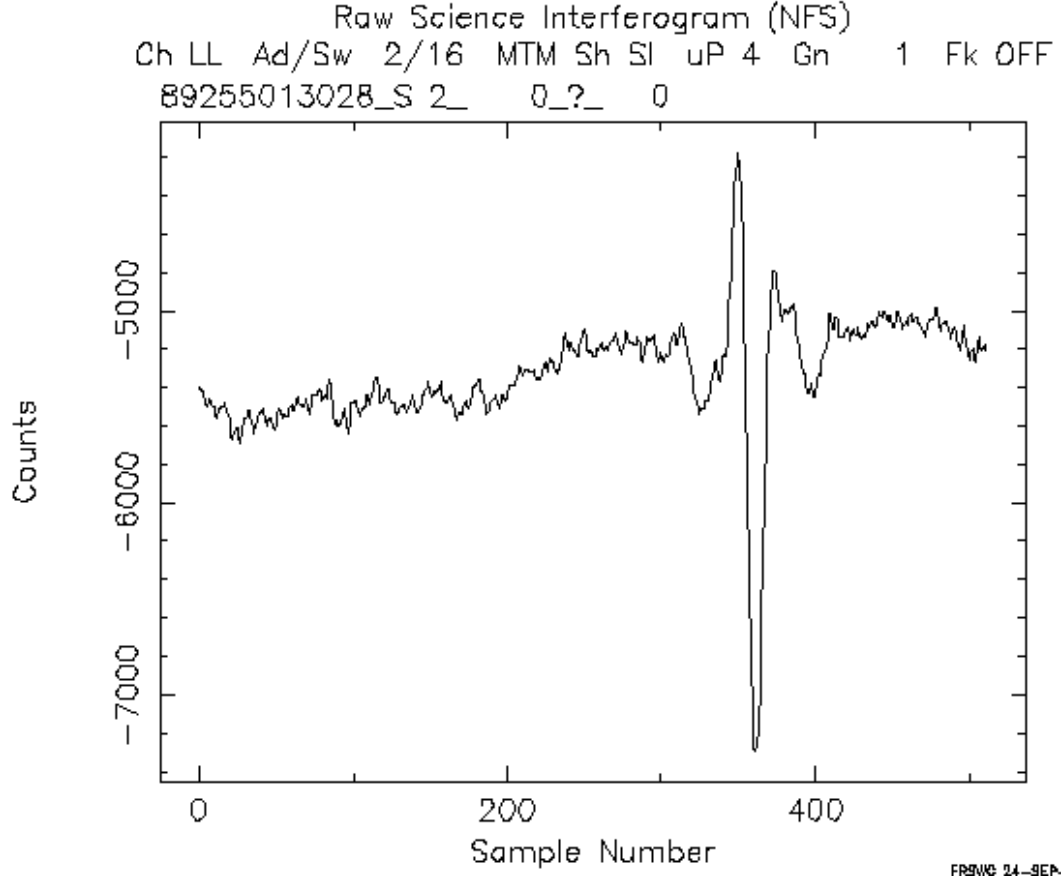}
\caption{A FIRAS interferogram from the Left Low channel
taken during ground test with the XCAL at 2.2~K and the ICAL at 2.7~K}
\label{ifg}
\end{figure}

Since the FIRAS is a Michelson interferometer, the spectral data is obtained 
in the form of interferograms.  Thus the LEFT output is approximately
\begin{equation}
I_L(x) = \int_0^\infty \cos(2\pi\nu x) 
G(\nu) \left(I_\nu + \sum_i \epsilon_i(\nu) B_\nu(T_i) + U_\nu \right) d\nu
\end{equation}
The index $i$ above runs over all the components in the FIRAS that had
thermometers to measure $T_i$.  These include the ICAL, the reference horn 
that connects to the ICAL, the sky horn, the bolometer housing, the optical
structure of the FIRAS, and the dihedral mirrors that move to provide the
variation in path length difference $x$.
The $\epsilon_i$'s are the effective emissivities of the various components.
The ICAL itself has $\epsilon \approx -1$, while the sky and reference horns
have $\epsilon$'s of $\pm$ a few percent.
The offset term $U_\nu$ was observed during flight to be approximately
$10^{-5}\exp(-t/\tau)B_\nu(T_U)$ with a time constant $\tau$ of two months
and a temperature $T_U \approx 15$~K.
The SKY input $I_\nu$ above can be either the sky or an external calibrator.
The XCAL is a movable re-entrant absorber that can be inserted at the top of
the sky horn.  The combination of sky horn plus XCAL forms a cavity with an
absorbtivity $> 0.999$, and we believe $\gsim 0.99999$.
With the XCAL inserted during periodic calibration runs, the SKY input is
known to be $B_\nu(T_X)$.  By varying $T_X$ and the other $T_i$'s, the
calibration coefficients have been determined.  See Fixsen \etal\ (1993a) for 
a more complete description of the calibration procedure.  Figure \ref{ifg}
shows an interferogram from ground testing with the ICAL at 2.7~K and the
XCAL at 2.2~K.

\begin{figure}[tb]
\plotone{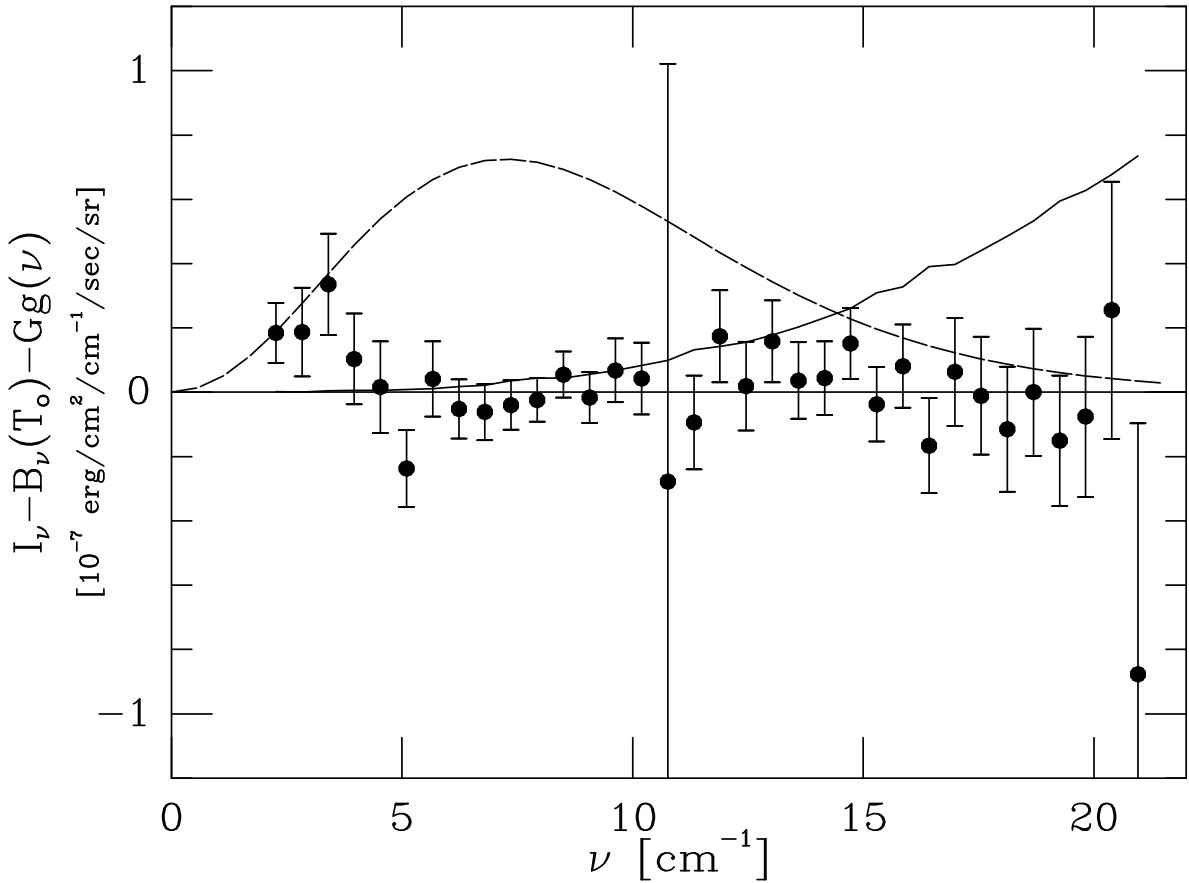}
\caption{FIRAS residuals (points) after fitting for a blackbody and
galactic emission, with an 0.5~mK $\Delta T$ and a
$\Delta(\csc|b|) = 0.25$ galaxy curve shown for comparison.  Full-scale
is 0.1\% of the peak of the CMB.}
\label{rtg}
\end{figure}

\begin{figure}[tb]
\plotone{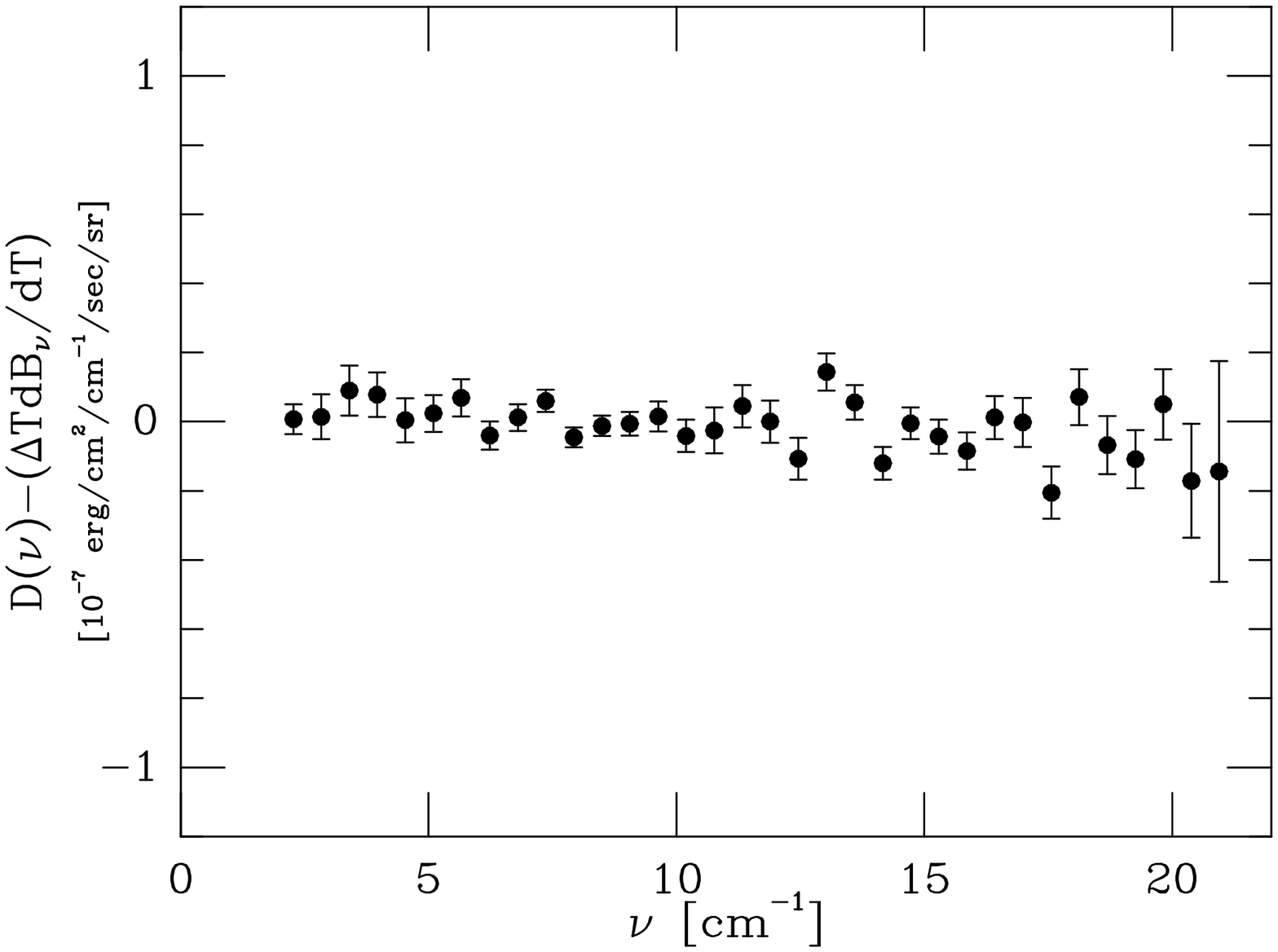}
\caption{FIRAS Dipole spectrum residuals (points).
Full-scale is 0.1\% of the peak of the CMB.}
\label{dipres}
\end{figure}

Once the calibration coefficients are known, the sky data can be analyzed to
determine $I_\nu(l,b)$ the intensity of the sky as a function of frequency,
galactic longitude and galactic latitude.  This is a ``data cube''.  The data
from each direction on the sky can be written as a combination of cosmic plus
galactic signals:
\begin{eqnarray}
I_\nu(l,b) & = & e^{-\tau_\nu(l,b,\infty)} 
\left(B_\nu\left(T_\circ + \Delta T(l,b)\right) 
+ \Delta I_\nu \right)\nonumber 
\\
& + & \int e^{-\tau_\nu(l,b,s)} j_\nu(l,b,s) ds
\end{eqnarray}
where $\tau_\nu(l,b,s)$ is the optical depth between the Solar system and 
the point at distance $s$ in the direction $(l,b)$ at frequency $\nu$,
$\Delta I_\nu$ is an isotropic cosmic distortion, and
$\Delta T(l,b)$ is the variation of the background temperature around its 
mean value $T_\circ$.  This equation can be simplified because the optical
depth of the galactic dust emission is always very small in the
millimeter and sub-millimeter bands covered by FIRAS.  
\begin{equation}
I_\nu(l,b)  \approx B_\nu\left(T_\circ + \Delta T(l,b)\right) + \Delta I_\nu
 +  \int j_\nu(l,b,s) ds
\end{equation}
Even in the optically
thin limit, some restrictive assumptions about the galactic emissivity
$j_\nu$ are needed, since the galactic intensity
$\int j_\nu(l,b,s) ds$ is a function of three variables, just like the
observed data.  The simplest reasonable model for the galactic emission
is the one used in Wright \etal\ (1991): let 
\begin{equation}
\int j_\nu(l,b,s) ds = G(l,b) g(\nu). 
\end{equation}
This model assumes that the shape of the galactic spectrum is independent
of direction on the sky.  It is reasonably successful except that the 
galactic center region is clearly hotter than the rest of the galaxy.
The application of this model proceeds in two steps.  The first step
assumes that the cosmic distortions vanish, and that an approximation
$g_\circ(\nu)$ to the galactic spectrum is known.  A least squares fit
over the spectrum in each pixel then gives the maps $\Delta T(l,b)$
and $G(l,b)$.  The high frequency channel of FIRAS is used to derive
$G(l,b)$ because the galactic emission is strongest there.  An alternative
way to derive $G(l,b)$ is to smooth the DIRBE map at 240 \um\ to the FIRAS
$7^\circ$ beam.  This DIRBE method has been used in the latest FIRAS spectral
results in Mather \etal\ (1993), Fixsen \etal\ (1993b) and 
Wright \etal\ (1993).  The second step in the galactic fitting then derives
spectra associated with the main components of the millimeter wave sky:
the isotropic cosmic background, the dipole anisotropy, and the galactic
emission.  This fit is done by fitting all the pixels (except for 
the galactic center region with $|b| < 20$ and $|l| < 40$)
at each frequency to the form
\begin{equation}
I_\nu(l,b) = I_\circ(\nu) + D(\nu)\cos\theta + G(l,b)g(\nu).
\end{equation}
The spectra of the anisotropic components derived from this fit applied
to the entire mission data set are shown in Figures \ref{rtg} and
\ref{dipres}.  Figure \ref{dipres} shows the residual after the
predicted dipole spectrum $\Delta T \partial B_\nu/\partial T$ evaluated
at $T_\circ$ is subtracted from $D(\nu)$.

But the FIRAS calibration model is so complicated and the cosmic distortions,
if any, are so small, that there are still systematic uncertainties that
limit the accuracy of the isotropic spectrum $I_\circ(\nu)$ derived from the
whole FIRAS data set.  Therefore the best estimate of $I_\circ(\nu)$ comes
from the last six weeks of the mission, when a modified observing sequence
consisting of 3.5 days of observing the sky with the ICAL set to null out
the cosmic spectrum, followed by 3.5 days of observing the XCAL with its
temperature set to match the sky temperature $T_\circ$.  Six cycles of
this alternation between sky and XCAL were obtained before the helium
ran out.  By processing both the sky data and the XCAL data through the
same calibration model, and then subtracting the two spectra,
almost all of the systematic errors cancel out.  The residual in the
high galactic latitude sky is then modeled as
\begin{equation}
I_\nu(sky,|b| > b_c) - I_\nu(XCAL) = \Delta I_\nu + 
\delta T \frac {\partial B_\nu} {\partial T} + G g(\nu).
\label{resid}
\end{equation}
The result of a least squares fit to minimize $\Delta I_\nu$ by
adjusting $\delta T$ and $G$ is shown in Figure \ref{rtg}.
The data points show the residuals while the curves show the
``uninteresting'' models $\partial B_\nu/\partial T$ and $g(\nu)$.
It is important to remember that any cosmic distortion with a spectral
shape that matches one of these curves will be hidden in this fit.
The maximum residual between 2 and 20 \percm\ is 1 part in 3000 of the peak
of the blackbody, and the weighted RMS residual is 1 part in $10^4$ of the
peak of the blackbody.
While these residuals are very small, they are nonetheless more than twice
the residuals that one would expect from detector noise alone.  The error
bars in Figure \ref{rtg} have been inflated by a constant factor to
give $\chi^2 = 32$ for the 32 degrees of freedom in the fit.
Limits on cosmological distortions can now be set by adding terms to the
above fit, finding the best fit value with minimum $\chi^2$, and then
finding the endpoints of the 95\% confidence interval where 
$\chi^2 = \chi^2_{min} + 4$.  
As an example of such a fit, suppose that the sky in fact was a gray-body
with an emissivity $\epsilon = 1+e$, where $e$ is a small parameter.
Since the XCAL is a good blackbody, this model predicts a residual of the
form $eB_\nu(T_\circ)$.  
Because $B_\nu(T_\circ)$ peaks at a lower frequency than
$\partial B_\nu/\partial T$, this model is sufficiently different from the
``uninteresting'' models and a tight limit on $e$ can be set:
$|e| < 0.00041$ (95\% confidence).
The result of fits for the usual cosmological suspects are
$|y| < 2.5 \times 10^{-5}$ and $|\mu| < 3.3 \times 10^{-4}$ with 95\%
confidence.

The absolute temperature of the cosmic background, $T_\circ$, can be
determined two ways using FIRAS.  The first way is to use the readings
of the germanium resistance thermometers in the XCAL when the XCAL
temperature is set to match the sky.  This gives $T_\circ = 2.730$~K.
The second way is to measure the frequency of the peak of 
$\partial B_\nu/\partial T$ by varying $T_X$ by small amounts
around the temperature which matches the sky, and then apply
the Wien displacement law to convert this frequency into a
temperature.  This calculation is done automatically by the
calibration software, and it gives a value of 2.722~K for
$T_\circ$.  The final adopted value is $2.726 \pm 0.010$~K
(95\% confidence),
which just splits the difference between the two methods.

\section{FIRAS Interpretation}

When interpreting the implications of the FIRAS spectrum on effects
that could distort the microwave background, it is important to
remember some basic order-of-magnitude facts.  Theoretical analyses
of light element abundances gives limits on the current density of
baryons through models of Big Bang NucleoSynthesis (BBNS).
Walker \etal\ (1991) give a ratio of baryons to photons of
$\eta = (3.5 \pm 0.7) \times 10^{-10}$, while Steigman and Tosi (1992)
give $\eta = (3.95 \pm 0.25) \times 10^{-10}$.  Since the number
density of photons is well determined by the FIRAS spectrum
\begin{equation}
N = 8 \pi \zeta(3)\Gamma(3) \left( \frac {k T_\circ} {h c} \right)^3
  = 410 \pm 5\;\percmcub
\end{equation}
the baryon density is well determined: $1.62 \times 10^{-7}\;\percmcub$.
This gives $\Omega_B h^2 = 0.0144$, where 
$h = H_\circ/(100\;{\rm km/sec/Mpc})$.
Not all of these baryons are protons, however.  For a primordial
helium abundance of 23.5\% by mass, one has 13 protons and 1 $\alpha$
particle in 17 baryons.  This gives an electron density of 
$n_{e,\circ} = 1.43 \times 10^{-7}\;\percmcub$.
For a fully ionized Universe, the optical depth for electron scattering
over a path length equal to the Hubble radius $c/H_\circ$ is $0.00088/h$.

The expansion of the Universe $H$ varies with time and redshift in a way
that depends on ratios of the current matter, radiation and vacuum densities
to the critical density $\rho_c = 3H^2/8\pi G$.  Let these ratios be
$\Omega_m$, $\Omega_r$, and $\Omega_v$.  Then
\begin{equation}
H(z) = (1+z) H_\circ \sqrt{(1-\Omega_m-\Omega_r-\Omega_v) + \Omega_m(1+z)
+ \Omega_r (1+z)^2 + \Omega_v(1+z)^{-2}}
\end{equation}
This equation is not exact for massive neutrinos which shift from being
part of $\Omega_m$ into part of $\Omega_r$ as $z$ increases.
The value of $\Omega_r h^2 = (4.16 \pm 0.06)\times 10^{-5}$ 
is fairly well-determined for the case of three massless neutrinos.
Thus if $\Omega_m \approx 1$ now,
the Universe becomes radiation dominated for 
$z > z_{eq} = 2.4h^2 \times 10^4$ when the background temperature was
$kT_{eq} = 5.65h^2\;{\rm eV}$.
In the radiation dominated epoch, $H = 0.645 (1+z)^2\;{\rm km/sec/Mpc}$ or
$H = 2.09(1+z)^2 \times 10^{-20}\;{\rm s}^{-1}$.

For early epochs when the Universe was ionized, electron scattering is
the dominant mechanism for transferring energy between the radiation field
and the matter.  As long as the electron temperature is less than
about $10^8$~K, the effect of electron scattering on the spectrum can
be calculated using the Kompaneets equation:
\begin{equation}
\frac {\partial n}{\partial y} = x^{-2} \frac{\partial}{\partial x}
\left[x^4\left(n + n^2 + \frac{\partial n}{\partial x}\right)\right]
\end{equation}
where $n$ is the number of photons per mode ($n = 1/(e^x-1)$ for a blackbody),
$x = h\nu/kT_e$, and the Kompaneets $y$ is defined by
\begin{equation}
dy = \frac{k T_e}{m_e c^2} n_e \sigma_T c dt.
\end{equation}
Thus $y$ is the electron scattering optical depth times the electron 
temperature in units of the electron rest mass.  Note that the electrons
are assumed to follow a Maxwellian distribution, but that the photon
spectrum is completely arbitrary.

Since the Kompaneets equation is describing electron scattering, which
preserves the number of photons, one finds that the $y$ derivative of the
photon density $N$ vanishes:
\begin{eqnarray}
\frac {\partial N}{\partial y} & \propto &\int x^2 \frac {\partial n}
{\partial y} dx \nonumber \\
 & = & \int \frac{\partial}{\partial x}
\left[x^4\left(n + n^2 + \frac{\partial n}{\partial x}\right)\right] dx 
\nonumber \\
 & = & 0 
\end{eqnarray}
The stationary solutions of the equation 
$\partial n/\partial y = 0$
are the photon distributions in thermal equilibrium with the electrons.
Since photons are conserved, the photon number density does not have to
agree with the photon number density in a blackbody at the electron
temperature.  Thus a more general Bose-Einstein thermal distribution
is allowed: $n = 1/(\exp(x+\mu)-1)$.  This gives
$\partial n/\partial y = 0$ for all $\mu$.
Since the Bose-Einstein spectrum is a stationary point of the Kompaneets
equation, it is the expected form for distortions produced at epochs
when
\begin{equation}
(1+z)\frac{\partial y}{\partial z} = 
\sigma_T n_{e,\circ} \frac {k T_\circ}{m_e c^2} \frac {c}{H}(1+z)^4 > 1
\end{equation}
For the $\Omega_B h^2$ given by BBNS, this redshift $z_y$ where
this inequality is crossed is well within the 
radiation dominated era, with a value
$z_y = 10^{5.1}/\sqrt{70\Omega_B h^2}$.

There are two simple solutions to the Kompaneets equation with non-zero
$\partial n/\partial y$ which both give the same spectral distortion.
The first simple case occurs when the electron temperature is much higher
than the typical photon energy.  This is the case for clusters of galaxies
today with $T_e \approx 10^8$~K, or for a hot intergalactic medium.
Since the variable $x$ in the Kompaneets equation is the photon energy
measured in units of the electron thermal energy, $x$ and hence $\Delta x$
are both very small in this case.  Therefore the $\partial n/\partial x$
term in the Kompaneets equation is much larger than the $n$ or $n^2$ terms
which may be dropped.  This gives a simpler equation
\begin{equation}
\frac {\partial n}{\partial y} = x^{-2} \frac{\partial}{\partial x}
\left[x^4\left(\frac{\partial n}{\partial x}\right)\right]
\end{equation}
which was used by Sunyaev and Zeldovich.  Note that the powers of $x$
on the right hand side cancel out, so the RHS can be evaluated using
$x = h\nu/k T_\gamma$ if the initial photon field is a blackbody with
temperature $T_\gamma$.  The solution is easily evaluated and usually 
expressed in terms of a frequency dependent brightness temperature:
\begin{equation}
\frac{\partial T}{\partial y} = T_\gamma 
\left(\frac {x (e^x+1)}{e^x-1} - 4 \right).
\end{equation}
The second simple case occurs when the initial photon field is a blackbody
with a temperature $T_\gamma$ which is only slightly below the electron 
temperature.  Letting $f = T_e/T_\gamma$, we find that the initial photon
field is given by $n = 1/(\exp(fx)-1)$.  Therefore
\begin{equation}
\left(n + n^2 + \frac{\partial n}{\partial x}\right) = 
\frac{(1-f)\exp(fx)}{(\exp(fx)-1)^2} 
= (1-f^{-1}) \frac{\partial n} {\partial x}.
\end{equation}
Thus the distortion has a Sunyaev-Zeldovich shape but is reduced in
magnitude by a factor $(1-T_\gamma/T_e)$.  Defining the ``distorting''
$y$ as
\begin{equation}
dy_D = \frac{k (T_e - T_\gamma)}{m_e c^2} n_e \sigma_T c dt
\end{equation}
we find that the final spectrum is given by a frequency-dependent
temperature given by
\begin{equation}
T_\nu = T_\circ \left[1 + y_D
\left(\frac {x (e^x+1)}{e^x-1} - 4 \right) + \ldots \right].
\end{equation}
where $x = h\nu/kT_\circ$.
The FIRAS spectrum in Figure \ref{rtg} shows that 
$|y_D| < 2.5 \times 10^{-5}$.

The energy density transferred from the hotter electrons to the cooler
photons in the $y$ distortion is easily computed.  The energy density is
given by $U \propto \int x^3 n dx$ so
\begin{equation}
\frac {\partial U}{\partial y_D} = 
\int x \frac {\partial}{\partial x}\left(x^4 \frac{\partial n}{\partial x}
\right) dx
\end{equation}
which when integrated by parts twice gives
\begin{eqnarray}
\frac {\partial U}{\partial y_D} & = &
-\int \left(x^4 \frac{\partial n}{\partial x} \right) dx \nonumber \\
 & = & 4 \int x^3 n dx = 4 U.
\end{eqnarray}
Thus the limit on $y_D$ gives a corresponding limit on energy transfer:
$\Delta U/U < 10^{-4}$.
Any energy which is transferred into the electrons at redshifts $z > 7$
where the Compton cooling time is less than the Hubble time will be
transferred into the photon field and produce a $y$ distortion.
Since there are $10^9$ times more photons that any other particles except
for the neutrinos, the specific heat of the photon gas is overwhelmingly 
dominant, and the electrons rapidly cool (in a Compton cooling time) back
into equilibrium with the photons.  The energy gained by the photons
is $\Delta U = 4 y a T_\gamma^4$ which must be equal to the energy
lost by the electrons and ions: $1.5 (n_e + n_i) k \Delta T_e$.  Since 
$y = \sigma_T n_e (k T_e/m_e c^2) c \Delta t$ we find the Compton cooling
time
\begin{equation}
t_C = \frac{1.5 (1+n_i/n_e) m_e c^2 } {4\sigma_T c U_{rad}}
    = \frac{7.4 \times 10^{19}\;{\rm sec}}{(1+z)^4}
\end{equation}
for an ion to electron ratio of $14/15$.
This becomes equal to the Hubble time 
$(3.08568\times 10^{17} {\rm sec})/(h(1+z)^{1.5})$ at $(1+z) = 9h^{0.4}$.

At redshifts $z > z_y = 10^5$, there will be enough electron scattering to
force the photons into a thermal distribution with a $\mu$ distortion
instead of a $y$ distortion.  However, the normal form for writing a $\mu$
distortion does not preserve the photon number density.  Thus we should 
combine the $\mu$ distortion with a temperature change to give an effect
that preserves photon number.  The photon number density change with $\mu$
is given by
\begin{eqnarray}
N  & \propto & \int {\frac {x^2 dx} {\exp(x+\mu)-1}} \nonumber \\
   & = & \sum_{k=1}^\infty e^{-k\mu} \int x^2 e^{-kx} dx \nonumber \\
   & = & 2 \sum_{k=1}^\infty {\frac {e^{-k\mu}}{k^3} } \nonumber \\
   & = & 2 \left(\zeta(3) - \mu \zeta(2) + \ldots \right).
\end{eqnarray}
A similar calculation for the energy density shows that
\begin{equation}
U \propto 6 \left(\zeta(4) - \mu \zeta(3) + \ldots \right).
\end{equation}
In order to maintain $N = const$, the temperature of the photon field
changes by an amount $\Delta T/T = \mu\zeta(2)/(3\zeta(3))$.
Therefore, the energy density change at constant $N$ is
\begin{equation}
\frac {\Delta U}{U} = 
\left(\frac {4\zeta(2)}{3\zeta(3)}-\frac{\zeta(3)}{\zeta(4)}\right) \mu
= 0.714\mu.
\end{equation}
Thus the FIRAS limit $|\mu| < 3.3 \times 10^{-4}$ implies 
$\Delta U/U < 2.4 \times 10^{-4}$.
The ``improved'' form of the $\mu$ distortion, with a $\Delta T$
added to keep $N$ constant, can be given as a frequency dependent 
brightness temperature:
\begin{equation}
T_\nu = T_\circ \left(1 + \mu 
\left[\frac{\zeta(2)}{3\zeta(3)} - x^{-1}\right] + \ldots \right).
\end{equation}
In this form it is clear that a $\mu$ distortion has a deficit of
low energy photons and a surplus of high energy photons with respect
to a blackbody.  In this it is like the $y$ distortion, but the
crossover frequency is lower.

Finally, at high enough redshift the process of double photon Compton
scattering becomes fast enough to produce the extra photons needed to
convert a distorted spectrum into a blackbody.  Whenever a photon with
frequency $\nu$ scatters off an electron, there is an impulse 
$\propto h\nu/c$
transferred to the electron.  This corresponds to an acceleration
$a \propto h\nu^2/(m_ec)$ for a time interval $\Delta t \propto 1/\nu$.
The energy radiated in new photons is thus $\propto e^2h^2\nu^3/(m_e^2c^5)$
which is $\propto \alpha h\nu (h\nu/(m_ec^2))^2$.  Since the rate 
of scatterings per photon is $\propto \Omega_B h^2 (1+z)^3$, 
the overall rate of new photon creation is $\propto \Omega_B h^2 (1+z)^5$ 
while the Hubble time is $\propto (1+z)^{-2}$.
Burigana, De Zotti \& Danese (1991) properly consider the interaction
of this photon creation process with the Kompaneets equation and
show that the redshift from which
$1/e$ of an initial distortion can survive is
\begin{equation}
z_{th} = \frac {4.24 \times 10^5}
{\left[\Omega_B h^2 \right]^{0.4}}
\end{equation}
which is $z_{th} = 2.3 \times 10^6$ for the BBNS value of $\Omega_B h^2$.

\section{DMR Observations}

The Differential Microwave Radiometers (DMR) experiment on \COBE\ is
designed to measure small temperature differences from place to place on the
sky.  The DMR consist of three separate units, 
one for each of the three frequencies of
31.5, 53 and 90 GHz.  The field of view of each unit consists of two beams
that are separated by a $60^\circ$ angle that is bisected by the spin axis.
Each beam has a $7^\circ$ FWHM.  The DMR is only sensitive to the
brightness difference between these two beams.  This differencing is
performed by a ferrite waveguide switch that connects the receiver input
to one horn and then the other at a rate of 100 cycles per second.  The 
signal then goes through a mixer, an IF amplifier and a video detector.
The output of the video detector is demodulated by a lock-in amplifier
synchronized to the input switch.  The difference signal that results
is telemetered to the ground every 0.5 seconds.
Each radiometer has two channels: A and B.  
In the case of the 31.5 GHz radiometer, 
the two channels use a single pair of horns in opposite senses of circular
polarization.  In the 53 and 90 GHz radiometers there are 4 horns, and all
observe the same sense of linear polarization.

\begin{figure}[tb]
\plotone{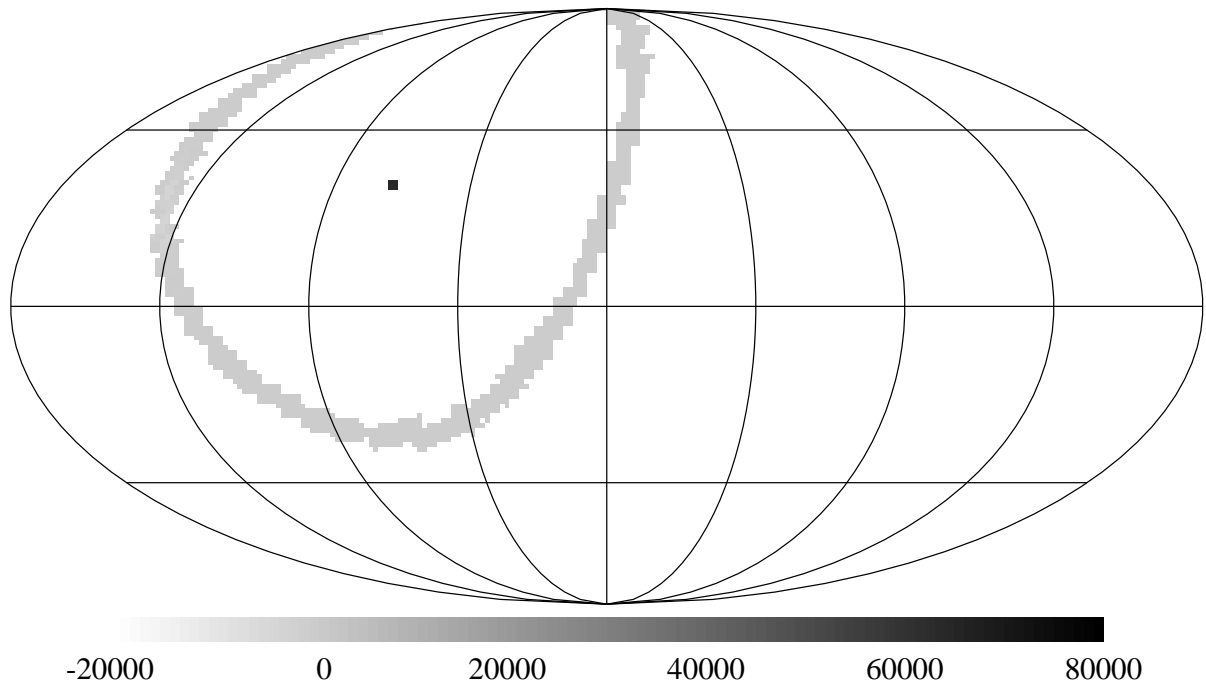}
\caption{The column through pixel 2719 of the sparse matrix A from a 2 year
simulation.}
\label{A2719}
\end{figure}

\begin{figure}[tb]
\plotone{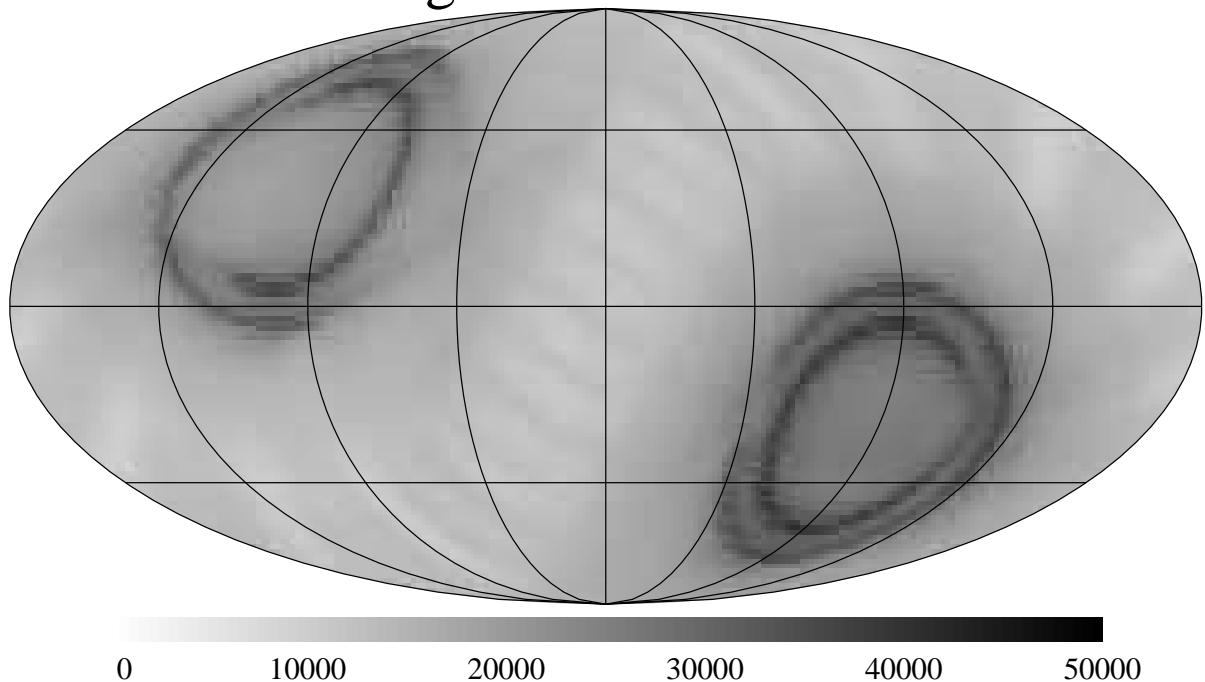}
\caption{The diagonal of the sparse matrix A, otherwise known as the
coverage map, from a 1 year simulation.}
\label{DIAG}
\end{figure}

The basic problem in the DMR data analysis is to construct a map of the
sky using the $6 \times 10^7$ differences that are collected each year.
There are only about $10^3$ beam areas on the sky, so the problem is
highly over-determined.  We have chosen to analyze the data using
6,144 pixels to cover the sky.  These are approximately equal area, and
arranged in a square grid of $32\times 32$ pixels on each of the 6 faces 
of a cube.  Within each pixel we assume that the temperature is constant,
so we are modeling the sky with a staircase function.  Therefore the
basic problem can be represented as a least-squares problem with 
$6\times 10^7$ equations in 6,144 variables.  Each equation has the form
$S_k = V_k X$ where the column vector $X$ is the map we wish to find, 
while the row vector $V_k = [0,\ldots,0,+1,0,\ldots,0,-1,0,\ldots]$
with the $+1$ in the pixel which contained the plus beam and the $-1$
in the pixel that contained the minus beam at the time of the $k^{th}$
observation $S_k$.  Then the normal least-squares equations are found
using
\begin{eqnarray}
A  & =  & \sum_k V_k^T V_k \nonumber \\
B  & =  & \sum_k S_k V_k^T \nonumber \\
X & = & lim_{\epsilon \rightarrow 0^+} \left(A + \epsilon I\right)^{-1}B
\end{eqnarray}
Note that $A$ is {\it sparse} and {\it symmetric}.
While $A$ has $38 \times 10^6$ elements, all but $1.8 \times 10^6$
of these elements are zero, and $A_{ij} = A_{ji}$.  
Thus only $9 \times 10^5$ elements need to be kept.

Visualizing a matrix with so many elements is difficult, but we can convert
a given row (or column by symmetry) into a map.  Figure \ref{A2719} shows
a column of the $A$ matrix.  The pixel on the diagonal is number 2719, and
it is one of the most heavily observed pixels in the rings $30^\circ$ from 
the ecliptic poles.  The diagonal of the $A$ matrix can also be made into
a map: it is the coverage map showing the number of observations in each
pixel (Figure \ref{DIAG}).

\begin{figure}[tb]
\plotone{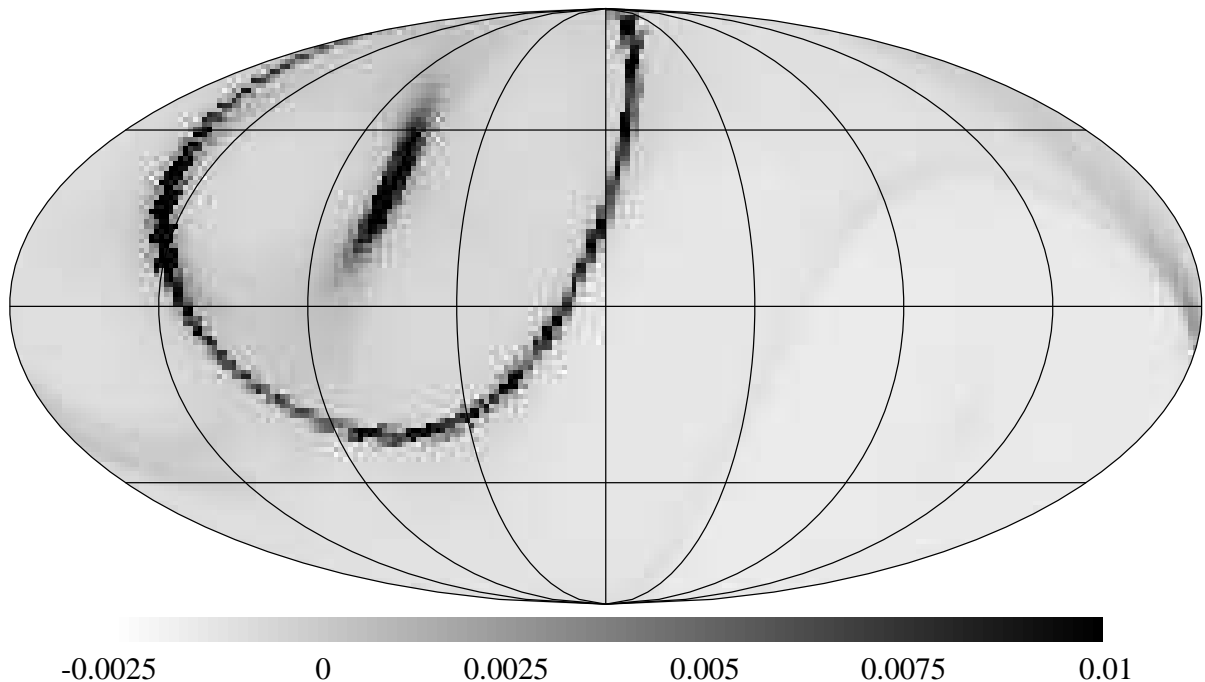}
\caption{The column through pixel 2719 of the normalized inverse of A.
The diagonal pixel 2719 is saturated with a value of 1.03}
\label{C2719}
\end{figure}

Adding $\epsilon I$ to $A$ allows for the fact that the sum or mean
of the map cannot be determined from the DMR data.  This is equivalent
to having one observation per pixel with weight $\epsilon$ which states
that the pixel value is zero.  Since there are $\approx 10^4$ observations
per pixel, with unit weight corresponding to an error of about 25~mK,
and the map values are all $< 4$~mK, rather large values
of $\epsilon \approx 10^3$ could be used without affecting the maps, while
the iterative algorithm used to find $X$ is quite stable even for 
$\epsilon \approx 10^{-3}$.  An alternative method that can be used
to regularize the inverse of $A$ is to add a matrix of all 1's to $A$.
This is equivalent to one observation stating that the sum of the map is
zero.  Both methods give the same answer for fully-covered sky maps,
but the $\epsilon I$ approach converges better for partial sky maps. 
Figure \ref{C2719} shows the result when $B$ above is replaced by the
column vector with elements 
$(-1,\ldots,-1,6143,-1,\ldots,-1)\times(N_{2719}/6144)$, where the positive
element is at pixel 2719, and $N_{2719}$ is the number of observations
at the $2719^{th}$ pixel.  The value of the central pixel
in this map is 1.03 which indicates that the conversion of the observed 
differences into a map value has increased the variance by 3\%.  The values
of about 0.01 or less in the $60^\circ$ radius reference ring indicates the
extent to which noise in the reference ring affects the map values.  These
pixel-pixel correlations are quite small and have very little effect on
any analyses except for the autocorrelation function of maps at $60^\circ$ 
separation.

\section{DMR Results}

Smoot \etal\ (1992), Wright \etal\ (1992), Bennett \etal\ (1992b) and
Kogut \etal\ (1992) announced the basic DMR result:  the discovery
of an intrinsic anisotropy of the microwave background, beyond the
dipole anisotropy discovered by Conklin (1969).  This anisotropy,
when a monopole and dipole fit to $|b| > 20^\circ$ is removed from
the map, and the map is then smoothed to a resolution of 
$\approx 10^\circ$, is 30 \muK.  The correlation function of this
anisotropy is well fit by the expected correlation function for
the Harrison-Zeldovich spectrum of primordial density perturbations
predicted by the inflationary scenario.
The ratio of the gravitational potential fluctuations seen by the DMR
through the Sachs-Wolfe (1967) effect to the gravitational potentials
inferred from the bulk flows of galaxies using the POTENT method
(Bertschinger \etal\ (1990)), shows that the Harrison-Zeldovich spectrum
has the correct slope to connect the DMR data at $3\times 10^5$~km/sec
scales to the bulk flow data at 6000~km/sec scale.

\begin{figure}[tb]
\plotone{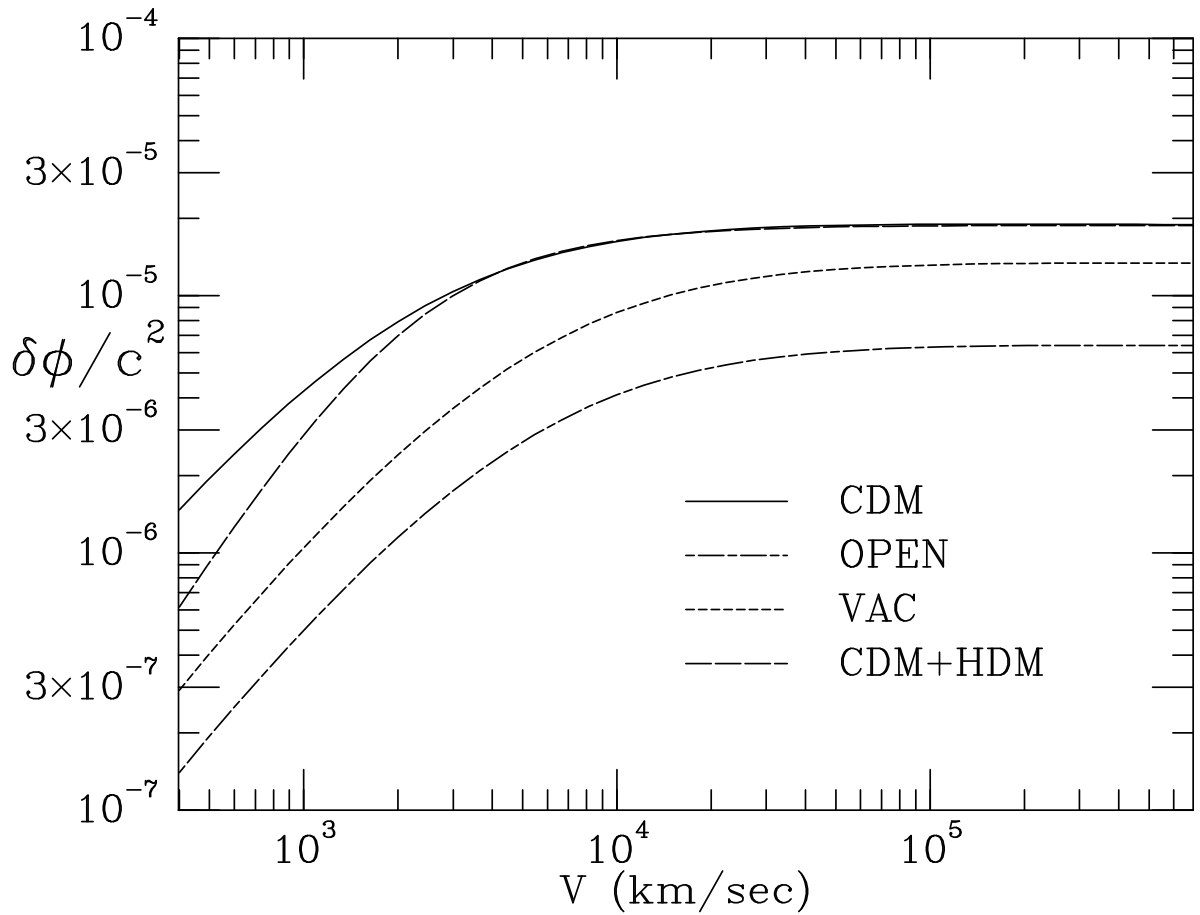}
\caption{The current gravitational potential perturbations
for the 4 models from Holtzman (1989) that were called out in
Wright \etal\ 1992.}
\label{pertpot}
\end{figure}

\begin{figure}[tb]
\plotone{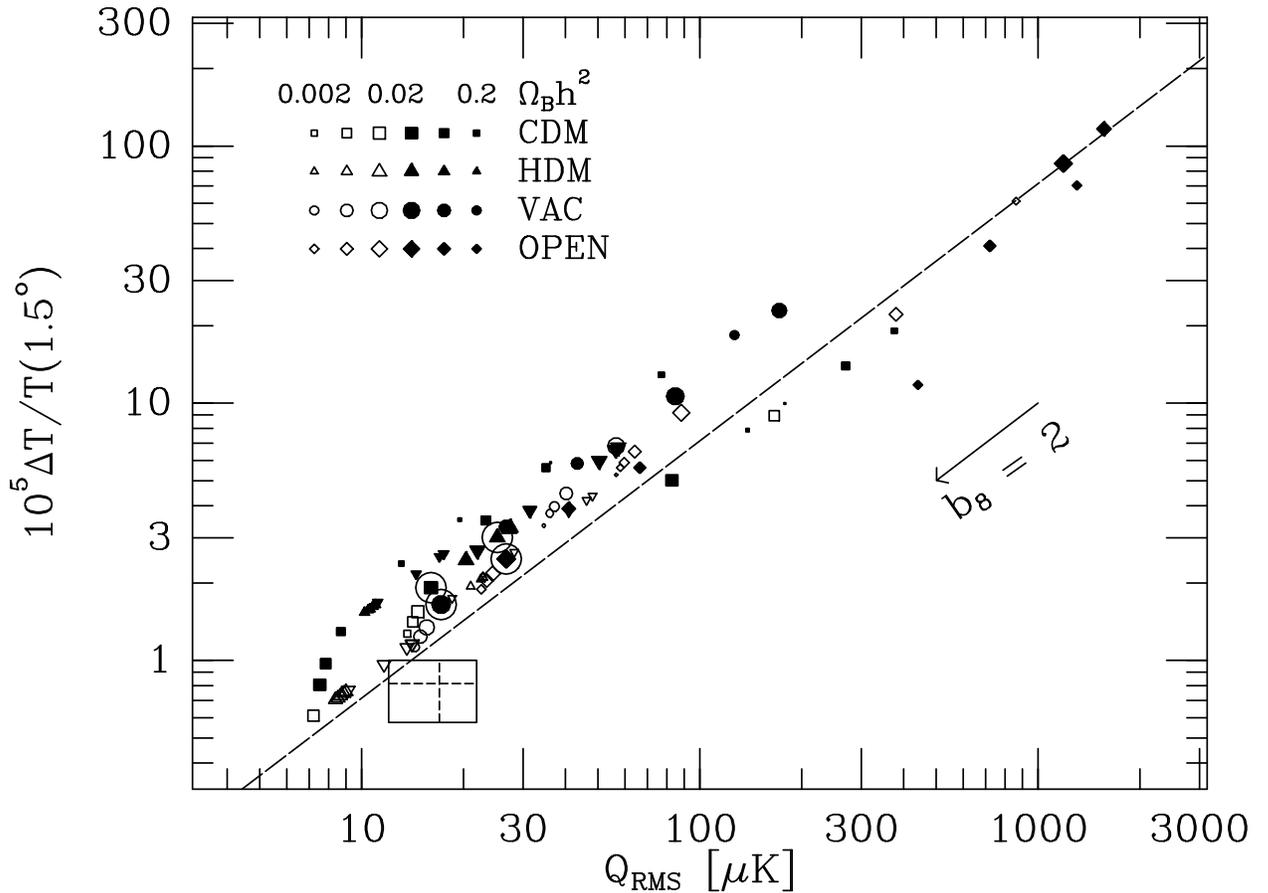}
\caption{The predicted anisotropy at the Gaier \etal\ (1992) scale
versus the predicted quadrupole for Holtzman models
compared to a pure Harrison-Zeldovich spectrum (diagonal line),
the Gaier \etal\ result reported at the 16$^{th}$ Texas symposium,
and the COBE \Amp\ from Smoot \etal\ (1992).  The ``$b_8 = 2$'' arrow
shows how the models move for a bias factor of 2 at $8/h$ Mpc.}
\label{gvq}
\end{figure}

Wright \etal\ (1992) discussed the implications of the \COBE\ DMR
data for models for structure formation, and selected 4 models
from the large collection in Holtzman (1989) for detailed discussion:
a ``CDM'' model, with $H_\circ = 50\;\kmsMpc$, $\Omega_{CDM} = 0.9$,
and $\Omega_B = 0.1$;
a mixed``CDM+HDM'' model, with $H_\circ = 50\;\kmsMpc$, $\Omega_{CDM} = 0.6$,
$\Omega_{HDM} = 0.3$ (a 7 eV neutrino), and $\Omega_B = 0.1$;
an open model, with $H_\circ = 100\;\kmsMpc$, $\Omega_{CDM} = 0.18$,
and $\Omega_B = 0.02$;
and
a vacuum dominated model, with $H_\circ = 100\;\kmsMpc$, 
$\Omega_{CDM} = 0.18$, $\Omega_B = 0.02$, and $\Omega_{vac} = 0.8$.
The vacuum dominated model and especially the open model
have potential perturbations now that are too small to explain the
Bertschinger \etal\ bulk flow observation, as seen in Figure \ref{pertpot}.
Note that all of these models are normalized to the \COBE\ anisotropy at
large scales at $z = 10^3$, but in the open and vacuum-dominated models
the potential perturbations go down even at very large scales which
are not affected by non-linearities or the transfer function.
The small-scale anisotropies predicted by these models are quite close
to the upper limits, and in the case of the $1.5^\circ$ beam South Pole
experiment they exceed the current data.  Figure \ref{gvq} shows the
status of this comparison.  Since the DMR datum is at $\ell_{eff} = 4$
while the Gaier \etal\ datum is at $\ell_{eff} = 44$, a tilted model
which replaces the Harrison-Zeldovich $n = 1$ in $P(k) \propto k^n$
with an $n \approx 0.5$ would eliminate this discrepancy.  But such a low
$n$ would destroy the agreement between the bulk flow velocities and
the COBE \Amp.  In fact the bulk flows are inconsistent with the Gaier
\etal\ data (Gorski 1992).

\section{Summary}

So far \COBE\ has been a remarkably successful space experiment with
dramatic observational consequences for cosmology.  The third instrument
on \COBE\ is the DIRBE instrument, which I have not had time to talk about.
The DIRBE is searching for a cosmic infrared background, and faces
the difficulty that the CIB is expected to be many times fainter than the
local infrared backgrounds from the Solar System and the Milky Way.

\section{References}

\bookref
Bennett, C. L. \etal\ 1992a.  COBE Preprint 92-08, ``Recent Results
from COBE'', to be published in The Third Teton Summer School: The
Evolution of Galaxies and Their Environment, eds. H. A Thronson \&
J. M. Shull.

\refitem
Bennett, C. L. \etal\ 1992b! ApJL! 396! L7;

\refitem
Bertschinger, E., Dekel, A., Faber, S. M., Dressler, A. \& Burstein, D. 1990!
ApJ! 364! 370-395;


\refitem
Burigana, C., De Zotti, G. F., \&  Danese, L. 1991! ApJ! 379! 1-5;

\refitem
Conklin, E. K. 1969! Nature! 222! 971;

\bookref
Fixsen \etal\ 1993a,  COBE Preprint 93-04, submitted to ApJ.

\bookref
Fixsen \etal\ 1993b,  COBE Preprint 93-02, submitted to ApJ.

\refitem
Gaier, T., Schuster, J., Gunderson, J., Koch, T., Meinhold, P., Seiffert, M.,
\& Lubin, P. 1992! ApJL! 398! L1;

\refitem
Gorski, K. 1992! ApJL! 398! L5-L8;

\refitem
Holtzman, Jon A. 1989! ApJSupp! 71! 1-24;

\refitem
Hubble, E. 1929! PNAS! 15! 168;

\refitem
Kogut, A. \etal\ 1992! ApJ! 401! 1-18;

\refitem
Martin, D. H. \& Puplett, E. 1970! Infrared Physics! 10! 105-109;

\bookref
Mather \etal\ 1993, COBE Preprint 93-01, submitted to ApJ.

\refitem
Penzias, A. A. \& Wilson, R. W. 1965! ApJ! 142! 419;

\refitem
Sachs, R. K. \& Wolfe, A. M. 1967! ApJ! 147! 73;

\refitem
Smoot, G. F. etal. 1992! ApJL! 396! L1;

\refitem
Steigman, G. \& Tosi, M. 1992! ApJ! 401! 150-156; 

\refitem
Walker, T. P., Steigman, G., Schramm, D. N., Olive, K. A. \& Kang, H-S. 1991!
ApJ! 376! 51-69;

\refitem
Wright, E. L. etal. 1992! ApJL! 396! L13;

\bookref
Wright \etal\ 1993,  COBE Preprint 93-03, submitted to ApJ.

\end{document}